\documentclass[fleqn,usenatbib]{mnras}

\usepackage{newtxtext,newtxmath}
\usepackage[T1]{fontenc}

\DeclareRobustCommand{\VAN}[3]{#2}
\let\VANthebibliography\thebibliography
\def\thebibliography{\DeclareRobustCommand{\VAN}[3]{##3}\VANthebibliography}

\usepackage{graphicx}	% Including figure files
\usepackage{amsmath}	% Advanced maths commands
\title[Super-competitive accretion]{The physical origin of super-competitive accretion during the formation of the first supermassive black holes}

\author[Schleicher et al.]{
Dominik R.G. Schleicher,$^{1}$\thanks{E-mail: dschleicher@astro-udec.cl}
Basti\'an Reinoso,$^{2}$
Ralf S. Klessen$^{2,3}$
\\
$^{1}$Departamento de Astronom\'ia, Facultad Ciencias F\'isicas y Matem\'aticas, Universidad de Concepci\'on, Av. Esteban Iturra s/n,Barrio Universitario, Concepci\'on, Chile\\
$^{2}$Universit\"at Heidelberg, Zentrum f\"ur Astronomie, Institut f\"ur theoretische Astrophysik, Albert-Ueberle Str. 2, 69120 Heidelberg, Germany\\
$^{3}$Universit\"at Heidelberg, Interdisziplin\"ares Zentrum f\"ur wissenschaftliches Rechnen, Im Neuenheimer Feld 205, 69120 Heidelberg, Germany
}

\date{Accepted XXX. Received YYY; in original form ZZZ}

\pubyear{2015}

\begin{document}
\label{firstpage}
\pagerange{\pageref{firstpage}--\pageref{lastpage}}
\maketitle

\begin{abstract}
Numerical simulations have shown the occurence of a scenario termed ''super-competitive accretion'', a term that describes a situation where only the central few objects grow supermassive while a larger number of stars compete for the reservoir, with significant accretion flows of  $\gtrsim0.1$~M$_\odot$~yr$^{-1}$. This scenario particularly implies that the presence of fragmentation will not necessarily impeed the formation of a central massive object.  We here explore this phenomenon using analytical estimates for growth via collisions and accretion, considering accretion due to self-gravity as well as Bondi-Hoyle accretion. Particularly, we explore under what conditions the accretion onto the central massive object breaks down, and derive a criterion that depends on the mass of the most massive object and the mass in fragments. For compact clusters with sizes about $0.1$~pc, we further find that the mass growth by collisions is comparable to the growth via accretion. Our results are validated through the comparison with numerical simulations, and we overall conclude that super-competitive accretion is a valid mechanism for the formation of very massive objects in the early Universe.
\end{abstract}


%%%%%%%%%%%%%%%%%%%%%%%%%%%%%%%%%%%%%%%%%%%%%%%%%%

%%%%%%%%%%%%%%%%% BODY OF PAPER %%%%%%%%%%%%%%%%%%

\section{Introduction}

Several different physical origins have been proposed for the first supermassive black holes that have been observed at high redshift \citep[e.g.][]{Fan06, Wu2015, Banados2018, Inayoshi2020}. A possible one is based on the remnants of the first stars, which have been proposed to be quite massive with $\sim100-1000~ $M$_\odot$ particularly in early investigations \citep[e.g.][]{Abel02, Bromm02}. More recent results indicate that, while very likely the masses are enhanced compared to present-day star formation due to the higher temperatures in the clouds, fragmentation is nevertheless expected to be relevant and should lead to lower masses by at least an order of magnitude \citep{Clark08, Clark2011, Greif2011, Susa2014, Riaz2018}. 

Another well-known scenario for the origin of supermassive black holes is based on direct collapse, i.e. the idea that a massive gas cloud will collapse into one single or at most of a few central objects. This idea has been examined analytically by \citet{Koushiappas2004} and \citet{Begelman2009} considering angular momentum transfer via gravitational torques. \citet{Bromm2003} explored this possibility via numerical simulations, showing the need to suppress molecular hydrogen cooling to form very massive central objects \citep[see also][for a discussion of relevant cooling mechanisms]{Schleicher2010}. Even in the very favorable case where cooling is only due to atomic hydrogen cooling, \citet{Latif2013} have shown that fragmentation cannot be fully avoided, but frequently at least two or three clumps will form on the characteristic timescale of the central accretion disk. However, suppressing the formation of molecular hydrogen is very difficult. {The first estimates based on numerical 3D simulations were provided by \citet{Shang2010}, and \citet{Wolcott2011} employed radiation-transport calculations to study the self-shielding of molecular hydrogen in protogalactic halos. \citet{Regan2014} have shown how the anisotropy of the radiation field in case of realistic radiation transport increases the required strength of the radiation field. }\citet{Sugirmura2014} and \citet{Agarwal2015} have shown via one-zone models that very strong external radiation fields are required when adopting realistic spectra to fully suppress H$_2$ formation. This effect is further enhanced in 3D simulations \citep{Latif2015}, where the virialization shock in the atomic cooling halo increases the abundance of free electrons and thus stimulates the H$_2$ formation.

In the presence of heavy elements or dust grains, fragmentation is expected to be further enhanced. This should lead to the formation of a dense cluster in the center of the halo \citep{Omukai2008}, where a massive black hole could form as a result of collisions between the protostars  \citep[see also semi-analytical work by][]{Devecchi2009}. This collision-based scenario was further explored in cosmological simulations by \citet{Katz2015} and via N-body simulations by \citet{Sakurai2017}, \citet{Reinoso2018, Reinoso2020} and \citet{Vergara2021}.

Of course, in realistic scenarios, black hole formation is unlikely to be only due to gas-dynamical or collision-based scenarios. On the one hand, gas-dynamical scenarios are likely to give rise to fragmentation and at least a temporary formation of clumps \citep[see e.g.][]{Grete2019, Suazo2019}. This by itself is not necessarily a problem, as gravitational torques can still lead to the merger of fragments with the central object {\citep{Inayoshi2014, LatifSchleicher2015}}. The high protostellar accretion rates in primordial gas typically also give rise to enhanced protostellar radii \citep[e.g.][]{Hosokawa2013, Schleicher2013, Haemmerle2018, Haemmerle2019, Haemmerle2021}, which will enhance the probability for collisions. \citet{Boekholt2018} explored the interplay of accretion and collisions using simplified models of the gas dynamics, showing that in such cases, central massive objects with up to $10^5$~M$_\odot$ could be produced. These simulations were extended by \citet{Seguel2020} considering the mass loss during collisions of protostars, showing that the latter leads to a maximum mass of the order $10^4-10^5$~M$_\odot$. A semi-analytic model for the formation of very massive objects via stellar bombardment and accretion was presented by \citet{Tagawa2020}. \citet{Das2021} pursued a systematic study to explore how different recipes for the accretion affect the probability of collisions, showing the collision probability to be enhanced in the presence of accretion, as the mass and the linear momentum of protostars are no longer conserved.

\begin{figure*}
    \centering
    \includegraphics[scale=0.7]{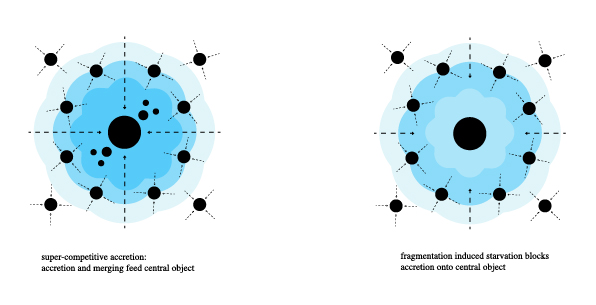}
    \caption{Illustration of super-competitive accretion (left) vs fragmentation-driven starvation (right). In the first case, accretion onto the central massive object continues in spite of the presence of other clumps. In the second case, the accretion onto such clumps becomes a limiting factor and prevents further growth of the central massive object.}
    \label{fig:illu}
\end{figure*}

Meanwhile, \citet{Chon2020} presented numerical simulations including dust cooling, in which fragments form, while the central massive object continues to accrete very efficiently; leading to a scenario termed "super-competitive accretion", where a few central objects grow supermassive while a larger number of fragments competes for the gas. The importance and potentially strong effect of gas inflows into nuclear star clusters has also been stressed by \citet{Lupi2014}, and more recently \citet{Kroupa2020}. Analytically, \citet{Schleicher2022} investigated the formation of massive objects in metal-poor clusters due to collisions and accretion, including the limitations due to radiative feedback. In principle, such scenarios need to be contrasted with the possibility of fragmentation-induced starvation \citep{Peters2010}, as in case of sufficient fragmentation around a massive object, the gas may be accreted onto the fragments and no longer reach the massive object. An illustration comparing the super-competitive accretion scenario and the fragmentation-induced starvation scenario is given in Fig.~\ref{fig:illu}.

The increasing importance of black hole formation scenarios where very massive objects form at very low metallicities, potentially due to a hybrid scenario based on accretion and collisions, have become apparent in particular through the statistical investigations by \citet{Sassano2021} and \citet{Trinca2022} with respect to the resulting black hole population. They find in particular that the formation of supermassive black holes becomes significantly more feasible if massive black hole seeds can form under conditions of very low but non-zero metallicity. In this range, we would normally expect fragmentation to occur, but the fragments may still merge and/or accrete fast, as proposed e.g. in the super-competitive accretion scenario. 

For these reasons, it is very important to understand further the conditions under which this phenomenon may occur. In this paper, we present an analytical assessment of the conditions that may give rise to super-competitive accretion  and assess also the potential for a significant growth via collisions. We will further review the possible contributions of accretion and collisions based on analytical estimates. Our analytical estimates are given in section \ref{analytical}. A quantitative comparison of these rates to distinguish the relevant regimes and compare with simulations from the literature is presented in section \ref{comparison}. Our main results are summarized and discussed in section \ref{summary}.

\section{Mass growth via collisions versus accretion}\label{analytical}

In this section, we provide the theoretical basis to analyze the growth of the central massive object via collisions and accretion. In subsection \ref{collisions}, we introduce the framework to estimate the mass growth via collisions. Subsection \ref{accretion} provides the formalism to assess growth via Bondi accretion and as a result of the self-gravity of the gas. Preliminary estimates on the expected effects and an assessment of the conditions that allow for super-competitive accretion is provided in subsection \ref{estimates}.

\subsection{Mass growth via collisions}\label{collisions}

Following \citet{Leigh2017}, the timescale for collisions with the most massive object can be evaluated as\begin{equation}
    t_{\rm coll}=\frac{G^3M^{1/2}\bar{m}M_{cl}^{13/2}}{12\sqrt{2}m_*^{3/2}N\left(R_{\rm MMO}+R_*\right )^2|E|^{7/2}},
\end{equation}
where $G$ is the gravitational constant, $M$ the mass of the most massive object, $\bar{m}$ the average mass of all particles including the most massive object, $M_{cl}$ the stellar mass of the cluster, $m_*$ the average mass of the stars (not considering the most massive object), $N$ the number of stars, $R_{\rm MMO}$ the radius of the most massive object, and $R_*$ the average stellar radius. $|E|$ refers to the total energy of the stars in the cluster, which we evaluate using the virial theorem as\begin{equation}
    |E|=\frac{GM_{\rm tot}M_{cl}}{2R_{cl}},
\end{equation}
with $R_{cl}$ the radius of the cluster and $M_{\rm tot}=M_{cl}+M_g$, with $M_g$ the gas mass of the cluster.

The mass growth via collisions can then be estimated as\begin{equation}
    \dot{M}_{\rm coll}=\frac{m_*}{t_{\rm coll}}.\label{coll}
\end{equation}

\subsection{Mass growth via accretion}\label{accretion}
The mass growth via accretion depends on the accretion rate. We will in the following consider accretion rates for different physical situations.

\subsubsection{Accretion due to self-gravity}
If the self-gravity of the gas dominates in the cluster, the corresponding critical mass scale, the Jeans mass, is given as \begin{equation}
    M_J=\left(\frac{4\pi \rho}{3}\right)^{-1/2}\left(\frac{5k_BT}{2\mu Gm_p}  \right)^{3/2},
\end{equation}
with $\rho$ the gas density, $k_B$ the Boltzmann constant, $T$ the temperature, $\mu$ the mean molecular weight and $m_p$ the proton mass. The free-fall time of the gas is given as\begin{equation}
    t_{ff}=\sqrt{\frac{3\pi}{32G\rho}}.
\end{equation}
The mass accretion rate if regulated via self-gravitating inflows is then given as\begin{equation}\label{BHsg}
    \dot{M}_{sg}=\frac{M_J}{t_{ff}}=\frac{2}{\pi}\sqrt{2G}\left(\frac{5k_BT}{2\mu Gm_p}  \right)^{3/2}.
\end{equation}

\subsubsection{Bondi-Hoyle accretion}
If the gravitational potential is dominated by the most massive object and the self-gravity of the gas can be neglected, the accretion rate is given by Bondi-Hoyle accretion as\begin{equation}
    \dot{M}_{BH}=\frac{4\pi G^2M^2\rho}{\left( V^2+c_s^2 \right)^{3/2}},\label{BH}
\end{equation}
where $V$ is the velocity of the most massive object relative to the gas, and $c_s$ the sound speed of the gas, which can be evaluated as\begin{equation}
    c_s=\sqrt{\frac{\gamma k_BT}{\mu m_p}},\label{cs}
\end{equation}
with $\gamma$ the adiabatic index of the gas. In the limit where the velocity of the most massive object can be negleected compared to the sound speed, we then obtain\begin{equation}
    \dot{M}_{BH}=4\pi G^2M^2\rho\left( \frac{\mu m_p}{\gamma k_BT} \right)^{3/2}.
\end{equation}

%\subsubsection{Eddington accretion}
%If the accretion rate is limited by radiative feedback under spherically symmetric conditions in an ionized plasma, it is given as\begin{equation}
 %   \dot{M}_{Edd}=\frac{4\pi GM}{\epsilon %\kappa c},
%\end{equation}
%with $\epsilon\sim0.1$ the radiative efficiency, $\kappa=0.4$~cm$^2$~g$^{-1}$ the electron scattering opacity and $c$ the speed of light.

%\subsubsection{Viscous accretion}
%If accretion is due to the viscosity $\nu$ in an accretion disk, it can be described as \citep{Latif2015} \begin{equation}
    %\dot{M}_{\rm vis}=3\pi\nu\Sigma, %\label{vis}
%\end{equation}

%with $\Sigma$ the surface density of the disk. For a thin disk we evaluate the viscosity following \citet{Shakura1973} %as\begin{equation}
%\nu=\alpha c_s H,\label{nu}
%\end{equation} 
%where $\alpha$ is a dimensionless parameter typically between $0.01$ and $1$, and $H$ is the scale height of the disk. 

\subsection{Preliminary estimates}\label{estimates}
We assess here the relative importance of the different accretion rates as well as concerning the relative importance of accretion and collisions.

\subsubsection{Self-gravity regulation versus Bondi-Hoyle accretion}
We first aim to understand when the accretion rate due to self-gravitating inflows dominates over the Bondi-Hoyle accretion rate. We assume that the central object is approximately at rest, i.e. $V\sim0$. Requiring that the accretion rate due to self-gravity, Eq.~\ref{BHsg}, is larger than the Bondi-Hoyle rate, Eq.~\ref{BH}, and expressing the sound speed via Eq.~\ref{cs} implies that
\begin{equation}
    M^2\rho<\frac{1}{2\sqrt{2}\pi^2G^3}\sqrt{5\gamma}^3\left( \frac{k_BT}{\mu m_p} \right)^3\sim1.1\times10^{44}g^3\ \mathrm{cm}^{-3}\left(\frac{\sqrt{\gamma} T/K}{\mu} \right)^3.
\end{equation}
In astrophysical units, this implies that
\begin{equation}
\left(\frac{M}{M_\odot}  \right)^2\left(\frac{\rho}{\mathrm{3.3\times10^{-18}\ g\ cm}^{-3}}\right)<\left( \frac{\sqrt{\gamma} }{\mu}\frac{T}{50\mathrm{\ K}} \right)^3.\label{astrosg}
\end{equation}

Provided that there is a sufficient gas reservoir, it is thus plausible that the accretion rate due to self-gravity will dominate at least in the beginning, when the density and the mass of the central object is sufficiently low. The temperature of the gas provides a relevant uncertainty, but even in the presence of efficient cooling via dust grains, a minimum temperature is provided by the cosmic microwave background, which we estimate to be of order $50$~K in the high-redshift Universe. This means that self-gravity may initially regulate the inflow of the gas even at densities of $\sim10^{-17}$~g~cm$^{-3}$, as long as the protostellar masses are less than a solar mass. In principle in the context of the formation of massive objects, it is then clear that accretion via self-gravity may occur at the earliest stages, while then subsequently Bondi-Hoyle accretion should be expected to take over. 

\subsubsection{Competition between different objects}

{From the Bondi-Hoyle estimate, one would thus expect Bondi-Hoyle accretion onto the central object to be the dominant process after an initial period. This could however} be avoided if the mass inflow to the central region would be preferentially accreted onto other objects, so that simply not even gas would be available for maintaing a Bondi-Hoyle rate. We assume here that we have $N$ fragments with an average mass $M_f$ and a velocity dispersion $v_f$, while the central object has mass $M$ and we assume we can neglect its motion relative to the gas. Accretion onto the central object will stop when all of the inflowing gas will be accreted by the fragments. Accretion onto the central object will definitely stop when the mass accretion onto the fragments will exceed the mass accretion onto the central object, so when\begin{equation}
\frac{NM_f^2}{v_f^3}=\frac{M_{cl}^2/N}{v_f^3}>\frac{M^2}{c_s^3},\label{crit}
\end{equation} 
where we used $M_f=M_{cl}/N$. We further assumed here that $v_f\gg c_s$, as in the limit of compact clusters, we expect the protostellar velocities to be comparable to the virial velocity of the cluster, while the sound speed will be regulated by cooling and thus in principle is not expected to increase for more compact configurations. Eq.~\ref{crit} translates in
\begin{equation}
M_{cl}>\sqrt{N\frac{v_f^3}{c_s^3}}M.\label{Nstop}
\end{equation}
While the velocity of the fragments could approximately correspond to the virial velocity in the cluster, the sound speed of the gas will usually be colder. If the temperature is regulated by atomic hydrogen cooling, we expect $T\sim8000$~K and thus $c_s\sim7.5$~km~s$^{-1}$ for a primordial gas. The virial velocity in the cluster can be estimated as $v_{\rm vir}\sim\sqrt{GM_{cl}/R_{cl}}$, where here we consider a compact cluster with $M_{cl}\sim3\cdot10^4$~M$_\odot$ and $R_{cl}\sim0.1$~pc, implying $v_f\sim v_{\rm vir}\sim36.5$~km~s$^{-1}$. We then have $\sqrt{v_f^3/c_s^3}\sim10.7$. In case of $N\sim100$, the criterion derived above then translates into $M_{cl}>100M$.

\begin{figure*}
    \centering
    \includegraphics[scale=0.38]{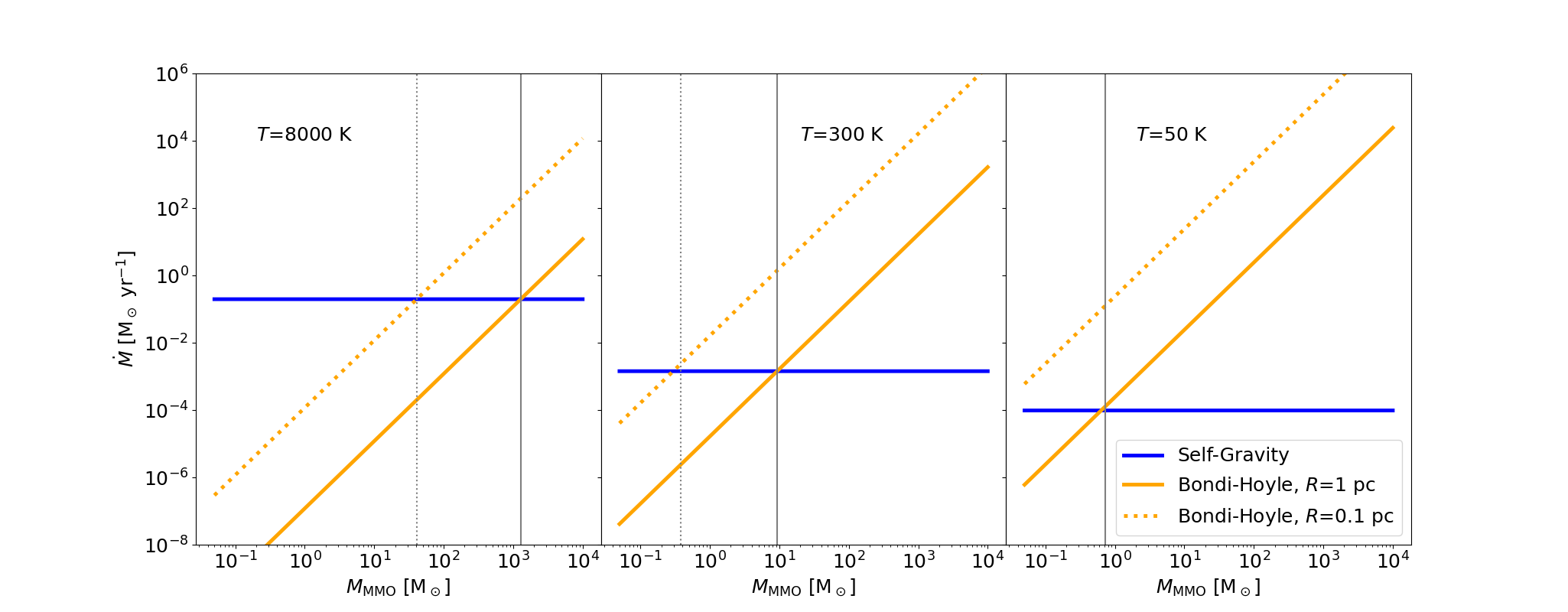}
    \caption{Mass growth rate as function of the mass of the MMO due to Bondi-Hoyle (yellow lines) and self-gravity regulated accretion (blue lines), for clusters of 1~pc (solid lines) and 0.1~pc in size (dotted lines). Each panel shows the rates calculated for different gas temperatures.}
    \label{fig:rate}
\end{figure*}

\begin{figure*}
    \centering
    \includegraphics[scale=0.38]{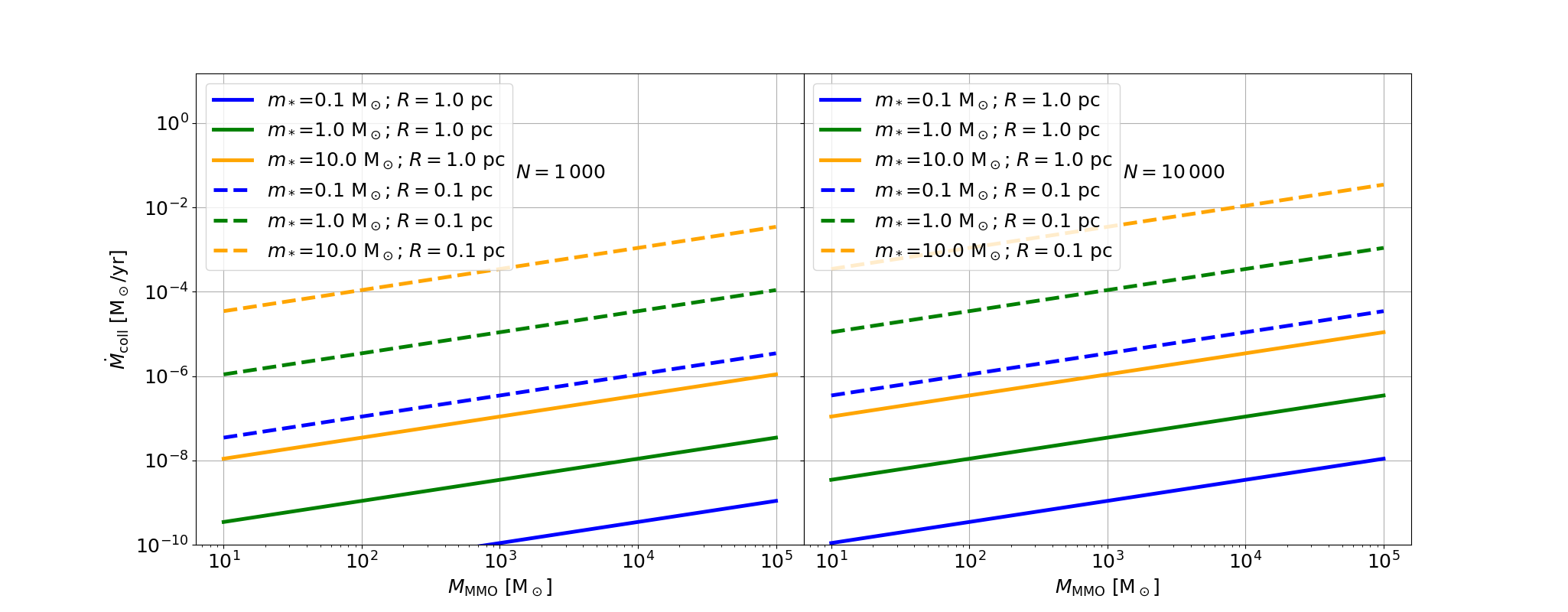}
    \caption{Mass growth rate as function of the mass of the MMO due to collision with surrounding fragments. The rate is calculated for clusters with 1.0~pc (solid lines) and 0.1~pc in size (dashed lines). We plot rates for fragment masses of 0.1~M$_\odot$ (blue lines), 1.0~M$_\odot$ (green lines), and 10.0~M$_\odot$ (yellow lines). Left panel shows the results for $N=1000$ fragments, whereas right panel shows the results for $N=10000$ fragments.}
    \label{fig:collgrowth}
\end{figure*}

\subsubsection{Influence of gravitationally bound objects within the Bondi radius}
As noted by \citet{Kaaz2019}, the Bondi accretion rate of the central object will be modified if other gravitationally bound objects are within its Bondi radius. Assuming that the $N$ fragments are uniformly distributed within a sphere of radius $R_{cl}$, we can compare their mean separation $R_{cl}/N^{1/3}$ to the Bondi-Hoyle radius of the central massive object,
given as\begin{equation}
        R_B=\frac{GM}{c_s^2}.
\end{equation}
The Bondi prescription should be valid as long as $R_{BH}<R_{cl}/N^{1/3}$, or\begin{equation}
%N<\left( \frac{M_g}{\rho} %\right)\frac{c_s^6}{GM^2},\label{NKaaz}
M<\frac{R_{cl}c_s^2}{GN^{1/3}}.\label{NKaaz}
%N<R^3\frac{c_s^6}{G^3M^3}.\label{NKaaz}
\end{equation} Both Eq.~\ref{Nstop} and Eq.~\ref{NKaaz} can be cast as a condition on the mass of the most massive object $M$, {where the first is a restriction due to the competition between different objects when accreting, while the second is a limitation in the Bondi-Hoyle approximation due to the presence of other objects within the Bondi-Hoyle radius. One}  may then ask which becomes relevant first. For Eq.~\ref{Nstop} to be more relevant, we need to require\begin{equation}
    \sqrt{\frac{c_s^3}{v_f^3N}}M_{cl}<\frac{Rc_s^2}{GN^{1/3}}
\end{equation}
or\begin{equation}
    M_{cl}<\frac{Rc_s^2}{G}\sqrt{\frac{v_f^3}{c_s^3}}N^{1/6}.
\end{equation}
We note in particular that this expression depends only weakly on the number of fragments, while the mass in protostars, the cluster radius, the sound speed and the ratio between sound speed and fragment velocity are much more significant. We express the equation here in astrophysical units:\begin{equation}
M_{cl}<1.4\times10^{16}M_\odot\left(\frac{R}{0.1\mathrm{\ pc}} \right)\left( \frac{c_s}{1\mathrm{\ km\ s^{-1}}}  \right)^2\left(  \frac{c_s/v_f}{0.01}  \right)^{3/2} \left(\frac{N}{100}  \right)^{1/6}.
\end{equation} 
We therefore conclude that Eq.~\ref{Nstop} will usually present the more relevant constraint.

\subsubsection{Limitations due to angular momentum}
Potentially, a limitation could arise from other factors. In principle the Bondi-Hoyle framework is only valid in the absence of angular momentum, while in the presence of rotation, a disk could form and accretion would proceed via viscous processes. As shown by \citet{Inayoshi2018}, such effects can be incorporated into the Bondi framework using a suppression factor\begin{equation}
    f_{\rm sup}=\left( \frac{\alpha}{0.01} \right)^{0.62}\left( \frac{r_{\rm in}}{R_{BH}} \right),
\end{equation}
where $r_{\rm in}$ refers to the inner radius of the accretion disk and they found a minimum value of $10^{-3}-10^{-2}$ for the suppression factor. In the presence of strong inflows, the viscous processes would likely be regulated via self-gravity, suggesting relatively high values of $\alpha$ between $0.1$ and $1$. A lower limit for the inner radius of the accretion disk can be obtained from the size of the protostar. In the presence of rapid accretion, it can be estimated as \citep{Schleicher2013}\begin{equation}
    R_*\sim2.6\times10^2R_\odot\left( \frac{M}{M_\odot}  \right)^{1/2},\label{radius}
\end{equation}
Usually this will still be considerably smaller than the Bondi radius, apart from late times when the central massive object reaches around $10^5$~M$_\odot$, when the two radii may become comparable. Here we will thus assume that the typical suppression factor will be in the range of the minimum suppression factor $10^{-3}-10^{-2}$ derived by \citet{Inayoshi2018}. In the competition of accretion between the central massive object and the fragment, this may shift the balance somewhat in favor of the fragments. As indicated above, the factor $\sqrt{Nv_f^3/c_s^3}$ in Eq.~\ref{Nstop} can be estimated to be of order $100$ for compact primordial clusters, so that the total stellar mass should exceed the mass of the most massive object by a significant degree in order to be a limiting factor.

\subsubsection{Growth via collisions}
We now estimate how the mass growth via collisions may evolve in comparison to the mass growth via accretion. For this purpose, we will simplify Eq.~\ref{coll} assuming $m_*\sim\bar{m}$, $R_*\ll R_{\rm MMO}$ and $R_{cl}\sim R$. We briefly note that this assumption is valid in the regime we discussed here, i.e. under the conditions of supercompetitive accretion. In general the mean mass of the protostars is given as\begin{equation}
\bar{m}=\frac{Nm_*+M}{N+1}\sim m_*+\frac{M}{N},\label{mean}
\end{equation}
where we assumed $N\gg1$, i.e. we assumed the number of fragments to be significantly larger than $1$. Now for the second term in Eq.~\ref{mean} to be comparable with the first, we would need that $Nm_*\sim M$. Our comparison with the literature below however shows that this is not the relevant regime for the formation of massive primordial objects. With the above assumptions, we then have\begin{equation}
\dot{M}_{\rm coll}\sim\frac{12\sqrt{2}R_{MMO}^2m_*^{3/2}N}{G^3M^{1/2}f_{\star}^3}\left( \frac{G}{2R} \right)^{7/2} \sqrt{M_{\rm tot}},
\label{coll_rate}
\end{equation}
with f$_{\star}$=M$_
{\rm cl}$/M$_{\rm tot}$.

Considering how $R_{\rm MMO}$ increases with the mass of the most massive object (Eq.~\ref{radius}), we find the accretion rate overall to increase with the central mass, but less steep compared to Bondi accretion.

%It is conceivable that the accretion rate onto the central object will not be given by the Bondi-Hoyle accretion rate. Particularly in the presence of rotation, a thin or thick disk may form and the accretion process will be governed by the transport of angular momentum within the disk. We estimate the disk radius as $R\sim(M/\rho)^{1/3}$, and parametrize the disk height as $H=hR$, with $h$ a dimensionless parameter characterizing the ratio of disk radius and disk height. Expressing $\Sigma\sim H\rho\sim hM^{1/3}\rho^{2/3} $, we have
%\begin{equation}
%\dot{M}_{\rm vis}\sim3\pi\alpha c_s h^2M^{2/3}\rho^{1/3}.
%\end{equation}
%Similarly as with Bondi-Hoyle accretion, a higher mass of the central object and a higher density thus enhances the accretion rate, but more moderately so. 

%\begin{equation}
%\sqrt{\frac{2G}{\pi^2}}\left(\frac{5}{\gamma G}  \right)^{3/2}>4\pi G^2M^2\rho
%\end{equation}

%subsection{Comparison of gas accretion rates}

%e should first check which processes are likely the most relevant ones in determining gas accretion. We have to determine a standard set of parameters, which could be $M_{cl}=M_g=10^5$~M$_\odot$, $R_{cl}=1$~pc. We can then compare the different accretion rates, considering the cases $M=100, 1000, 10^4$~M$_\odot$, $T=10, 100, 10^4$~K. This should (hopefully) give us an idea on which accretion rates are most likely to dominate in different physical regimes

\section{Comparison of accretion and collision rates}\label{comparison}
To assess the impact of the previously derived expressions, we provide a comparison of the analytical growth rates of the central object in subsection \ref{companal}, and a comparison with simulations from the literature in \ref{complit}.

\subsection{Comparison of the analytical growth rates}\label{companal}

%Once we have an idea which gas accretion rates are likely most relevant, we can also compare them with the expected mass growth rate through collisions. This should help us to understand when the accretion of the gas should be the dominant process.

We begin with a comparison of Bondi-Hoyle accretion and self-gravity regulated accretion. The mass in gas is equal to $10^5$~M$_\odot$. We study three different gas temperatures of 8000~K, 300~K, and 50 K, as well as clusters with a size of $R=1.0$~pc and $R=0.1$~pc. For the gas, we adopt $\gamma=1$ and $\mu=1.76$. For the protostars, we assume a uniform distribution where we will vary the number of protostars and their masses below. We also assume that the MMO is static and resides in the cluster center.  The cluster of protostars is embedded in a gas cloud following a Plummer density distribution, and estimate the density around the MMO as the mean gas density inside the central 1000~au of the cluster. We plot in Fig.~\ref{fig:rate} the mass growth rate of the MMO both via self-gravity from Eq.~\ref{BHsg} and Bondi-Hoyle mediated accretion from Eq.~\ref{BH}. 

Our results show that self-gravity regulated accretion is dominant at higher gas temperatures, in particular we find an accretion rate of 0.2~M$_\odot$~yr$^{-1}$ for $T=8000$~K. For lower gas temperatures of 300~K, self-gravity would dominate the accretion rate until the object gets to 0.4~M$_\odot$ if the cluster is 0.1~pc, or when the object reaches 10~M$_\odot$ if the cluster size is 1.0~pc. The accretion rate due to self gravity is of the order of 10$^{-3}$~M$_\odot$~yr$^{-1}$ in this case. Finally, for the lowest gas temperature of 50~K considered here, Bondy-Hoyle accretion always dominates in clusters of 0.1~pc, and dominates once the objects reaches 0.7~M$_\odot$ in clusters of 1.0~pc. The accretion rate due to self-gravity in this case is of the order of 10$^{-4}$~M$_\odot$~yr$^{-1}$.

For studying the growth rate through stellar collisions we investigate its dependence on the cluster size, the mass of the fragments, and on the number of fragments. We plot in Fig.~\ref{fig:collgrowth} the mass growth rate for clusters with $R=1.0$~pc (solid lines) and clusters with $R=0.1$~pc (dashed lines). For both models we also vary the mass of the fragments $m_*$ and plot the cases for 0.1~M$_\odot$ (blue lines), 1.0~M$_\odot$ (yellow lines), and 10~M$_\odot$ (green lines). We calculate the mass growth rate with Eq.~\ref{coll_rate}. We assume the same mass in gas and stars by setting f$_\star=1.0$ and use M$_{\rm cl}=10^5$~M$_\odot$. The radius of the MMO depends on its mass as indicated by Eq.\ref{radius}. We also vary the number of fragments $N$ and plot the rates for $N=1000$ on the left panel and $N=10000$ on the right panel.

\begin{figure*}
    \centering
    \includegraphics[scale=0.38]{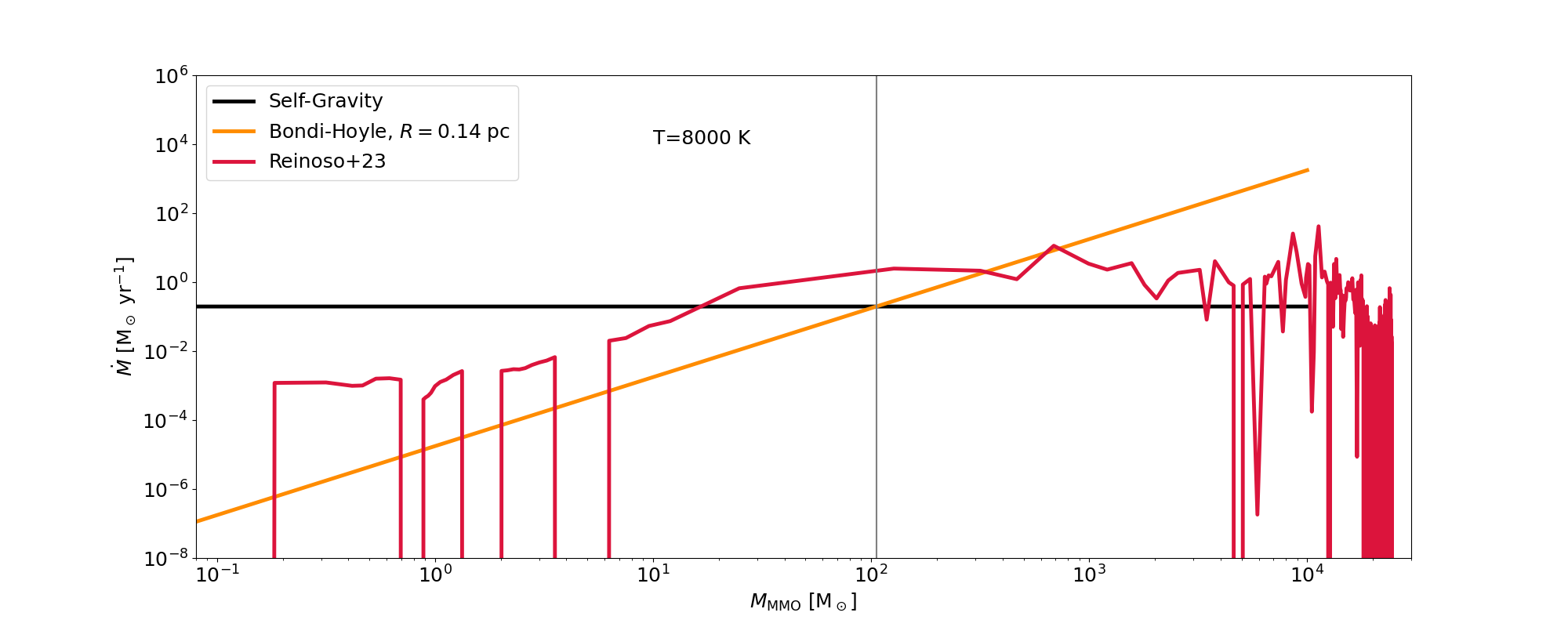}
    \caption{Mass growth rate as function of the mass of the MMO due to gas accretion through self-gravity and Bondi-Hoyle, along with accretion rates taken from \citet{Reinoso2023}.}
    \label{fig:accgrowthReinoso}
\end{figure*}

\begin{figure*}
    \centering
    \includegraphics[scale=0.38]{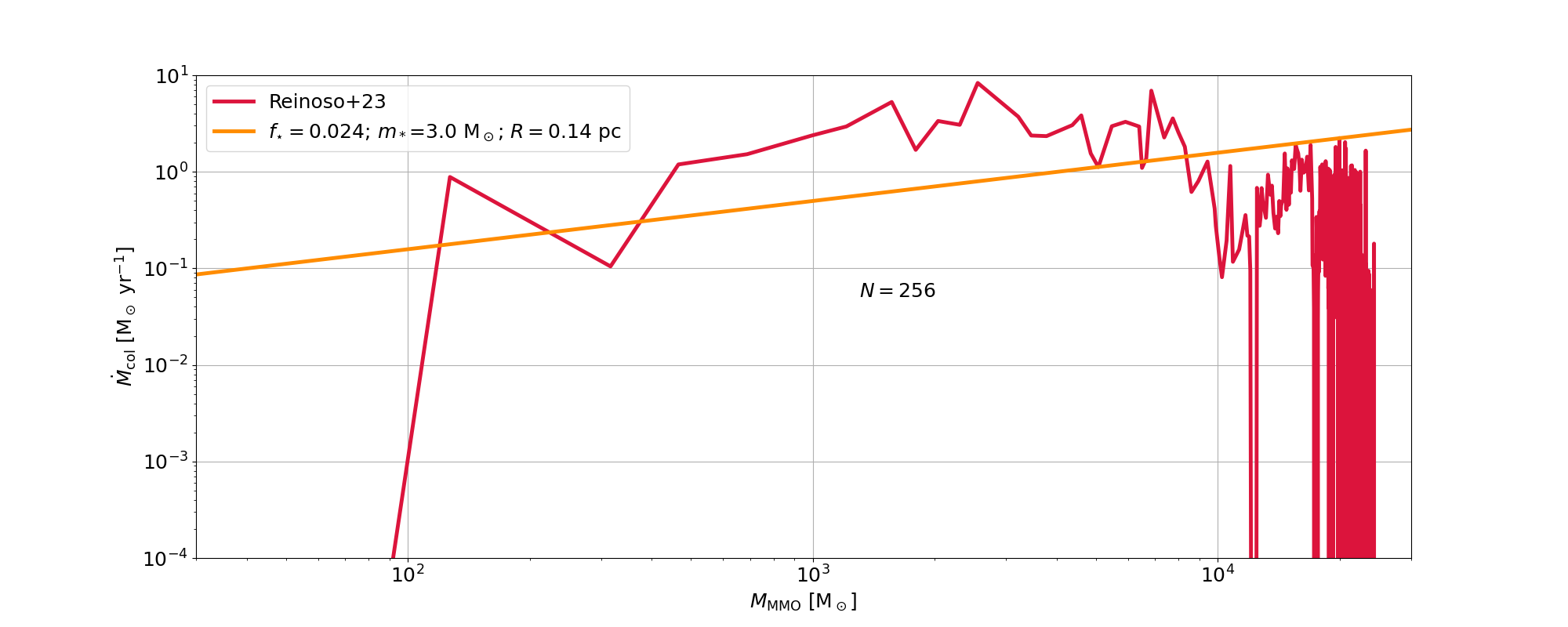}
    \caption{Mass growth rate due to collisions with fragments surrounding the MMO as calculated with the analytic rate in Eq.~(\ref{coll_rate}) along with results from the numerical simulations presented in \citet{Reinoso2023}.}
    \label{fig:collgrowthReinoso}
\end{figure*}

\subsection{Comparison with the literature}\label{complit}
An important point of comparison are the simulations by \citet{Chon2020}, as these originally introduced the concept of super-competitive accretion. We recall that they modeled massive gas clouds of $2\times10^4$~M$_\odot$, in a radius of $10^5$~au or $0.48$~pc, considering metallicities of $Z/Z_\odot=10^{-6}, 5\times10^{-6}, 10^{-5}, 10^{-4}$ and $10^{-3}$. The first metallicity corresponds to an effectively primordial regime with high temperatures of at least $3000$~K; the subsequent three cases include relevant amounts of dust cooling, with minimum temperatures in the range of $200-1000$~K, while the last case includes both dust and metal line cooling, with a minimum temperature of about $30$~K. In particular, it is notable that in the last simulation, significant cooling occurs already at densities of the order $3\times10^4$~cm$^{-3}$, while in the other simulations the pronounced cooling effects occur from densities of $10^{10}$~cm$^{-3}$ or higher. For a primordial metallicity, they find that only a small number of fragments forms, with the mass of the primary star reaching $\sim10^4$~M$_\odot$ within about $10^4$~yr. For most of the simulations, they find that an accretion rate onto the primary star in the range of $1-10$~M$_\odot$~yr$^{-1}$ can be maintained, while in the highest metallicity case, the accretion rate is about two orders of magnitude lower. The total stellar mass follows a similar trend and reaches more than $10^4$~M$_\odot$ for most simulations and about $10^3$~M$_\odot$ for the highest-metallicity case.  

From the simulations, it is thus notable that accretion onto the central object always occurs; though it is reduced if the temperature on the larger scales is being reduced as a result of metal-line cooling. Assuming a temperature on large-scales of order $8000$~K, we can estimate the inflow rate via Eq.~\ref{BHsg} obtaining about $0.2$~M$_\odot$~yr$^{-1}$. If on the other hand the temperature is reduced to about $400$~K due to metal line cooling on large scales, our estimate becomes $0.002$~M$_\odot$~yr$^{-1}$. The estimates are consistent with the trend we see in the simulations but somewhat lower, which we attribute to the fact that the modeled cloud indeed not consists of only one collapsing Jeans mass, but several Jeans masses, therefore leading to a larger accretion rate onto the central object. \citet{Chon2020} also include the growth of the central object via mergers both with lower-mass and higher mass stars. At metallicities of $\sim10^{-6}$~Z$_\odot$, they find the total contribution via mergers to be about $40\%$, at metallicities of $\sim10^{-3}$~Z$_\odot$ it becomes $70\%$.

\citet{Reinoso2023} have simulated protostellar clusters embedded in extremely metal poor gas clouds, considering a gas mass of $3\times10^4$~M$_\odot$ with a cloud half-mass radius of $0.1$~pc. Embedded in the cloud were 256 protostars with an initial mass of $0.1$~M$_\odot$. The gas temperature was isothermal at $8000$~K, corresponding essentially to a primordial gas where the molecular hydrogen is fully dissociated either due to collisions or a soft UV radiation field. After $200.000$~years, they found that about $80\%$ of the mass went into the central object, with accretion and collisions contributing about equally to the mass growth. In Fig.~\ref{fig:accgrowthReinoso}, we compare the accretion rate onto the most massive object obtained in their reference run with our analytical predictions both for the Bondi-Hoyle accretion rate and the accretion rate if the flow is driven by self-gravity, both as a function of the mass of the central object. For calculating the Bondi-Hoyle rate we take the mean gas density inside the inner 1000~au of a Plummer sphere with a virial radius of 0.14~pc. We assume the gas to be isothermal at 8000~K.
The accretion rate due to self-gravity overestimates the real accretion flow in particular at early stages, as the gas still needs to collapse and fall towards the center. The Bondi-Hoyle rate on the other hand somewhat underestimates the accretion flow when the mass of the central object is still low, most likely as it does not take the effect of self-gravity into account. At late times we find that the accretion rate expected due to self-gravity provides a good match for the mean rate. We also note in general that the accretion rate in the simulation is variable, due to fluctuations and substructure within the gas.

We also compare the growth rate due to stellar collisions from Eq.~\ref{coll_rate} with the simulations presented in \citet{Reinoso2023}. From the simulated data we find that most of the fragments are distributed around the MMO in a radius of $R=0.1$~pc inside which encloses an average total mass of $M_{\rm tot}=22500$~M$_\odot$ during the growth of the MMO. The fraction of that mass in fragments is on average $f_{\star}=0.024$, and we take the mean mass of the fragments to be $m_{\star}=3$~M$_\odot$, the typical mass of protostars that merge with the MMO in the simulations. Considering these data, we find a good agreement between the mass growth rate due to collisions in the simulation with the analytic estimate from Eq.~\ref{coll_rate} in Fig.~\ref{fig:collgrowthReinoso}. The growth rate due to collisions and accretion in the simulations is comparable. We note that the analytic rate underestimates the simulated growth rate during the initial growth of the MMO and then slightly overestimates it when the MMO grows to $\sim10^2$~M$_\odot$. This is likely due to the fact that some assumptions are not valid, in particular the distribution of fragments around the MMO is not uniform and the size varies. The gas distribution also presents substructures as it accumulates in a disk around the MMO. Furthermore the fragments have a wide range of masses, up to 100~M$_\odot$ and the number of fragments is not fixed but decreases as the MMO is close to $10^4$~M$_\odot$ when it has accreted most of the gas mass. Nevertheless the analytic rate provides a good approximation to estimate the mass obtained during the run.

A comparison with the simulations of massive star formation in present-day clouds \citep[e.g.][]{Peters2010} shows a different picture. They simulated star formation in molecular cloud cores of $1000$~M$_\odot$ within a radius of $0.5$~pc, where they compared simulations with and without radiative feedback. In both of them, the total mass in sinks was considerably higher than the mass of the most massive object, corresponding to $125.56$~M$_\odot$ in sinks and $23.39$~M$_\odot$ in the most massive object for the case with feedback, and  $151.43$~M$_\odot$ in fragments and $14.64$~M$_\odot$ in the most massive object without feedback. In this case it is thus clear that the fragmentation-induced starvation scenario is much more viable due to the larger amount of mass present in fragments.

\section{Summary and discussion}\label{summary}
In this paper, we have explored the physical origin of super-competitive accretion, as originally observed by \citet{Chon2020} in numerical simulations that included the effect of dust cooling at very low metallicities. In their work, they found that the central massive object continues to accrete at very high rates in spite of the occurence of fragmentation in the environment. The presence of such hybrid formation mechanisms of massive objects, including collisions and accretion, has been suggested more broadly in various studies, including \citet{Boekholt2018}, {\citet{Regan2020}}, \citet{Seguel2020}, \citet{Tagawa2020} and \citet{Schleicher2022}. Here our goal was to understand more fundamentally on analytical grounds in which cases efficient accretion onto a central object can continue, in spite of accretion. We further compared the contribution to the growth of the most massive object via accretion and collisions.

We provided an assessment of the dominant accretion mechanism, including accretion driven by the effects of self-gravity as well as Bondi-Hoyle accretion, where the letter includes corrections both for the velocity of the massive objects itself and potentially for the effects of angular momentum due to a suppression factor \citep[see][]{Inayoshi2018}. From a quantitative comparison, we found that the Bondi-Hoyle accretion rate typically starts to get more relevant for masses of the massive object between $1$ and $30$~M$_\odot$, depending on the temperature. The mass growth via collisions is calculated following \citet{Leigh2017} using the mean-free path approximation. Our estimates show that the growth via collisions becomes comparable to the growth via accretion for more compact clusters with typical radii of about $0.1$~pc.

To understand when mass growth via accretion will be impeded due to the presence of additional fragments \citep[fragmentation-induced starvation, see][]{Peters2010}, we have compared the expected accretion rate for the central object with the accumulated accretion rate onto the different fragments, to determine at which point the central object will run out of gas due to the accretion of the fragments. Additionally, the accretion onto the central object will become limited when some of the fragments lie within its Bondi radius, thereby reducing the available material that can be accreted. We have derived quantitative expressions for both of these limiting conditions. We found the first of these to be the more relevant one, which depends on the ratio between the mass in fragments and the mass in the massive object.

For the validation of the analytical models, we have performed comparisons with the simulations by \citet{Chon2020} and \citet{Reinoso2023}, finding a good general agreement in the expected behaviour. We note that the accretion rates in the simulations are sometimes still even higher, as in the analytical models (in case of accretion by self-gravity), the accretion rate was estimated based on the collapse of one Jeans mass, while the cloud modeled in the simulation consisted of several Jeans masses. The simulations are probing different regimes, as the ones by \citet{Chon2020} probe the regime where dust cooling becomes relevant at very low metallicities, while the \citet{Reinoso2023} simulations concern the regime of an atomic primordial gas.  An additional comparison with the results of massive star formation simulations in present-day clouds further shows that in the latter case, the total mass in sinks is much higher, so that fragmentation-induced starvation here becomes a more limiting scenario \citep{Peters2010}. {For a compact cluster of $3\times10^4$~M$_\odot$ and radius $0.1$~pc in the presence of 100 protostars, this condition translates in the protostellar mass being more than a hundred times the mass of the most massive object in order to produce fragmentation-induced starvation.}

Overall, we conclude here that fragmentation-induced starvation requires that the total mass in fragments becomes sufficiently high compared to the mass of the central massive object, roughly by a factor of $100$ for compact primordial clusters. The simulations in the literature however suggest that the protostellar masses do not reach this limit, which is the reason that accretion onto the central object continues to be very efficient. Overall, we conclude that supercompetitive accretion is a viable mechanism to form very massive objects under early Universe conditions.

\section*{Acknowledgements}
BR acknowledges support through ANID (CONICYT-PFCHA/Doctorado acuerdo bilateral DAAD/62180013) as well as support from DAAD (funding program number 57451854). The team acknowledges funding from the European Research Council via the ERC Synergy Grant ``ECOGAL'' (project ID 855130), from the Deutsche Forschungsgemeinschaft (DFG) via the Collaborative Research Center ``The Milky Way System''  (SFB 881 -- funding ID 138713538 -- subprojects A1, B1, B2 and B8) and from the Heidelberg Cluster of Excellence (EXC 2181 - 390900948) ``STRUCTURES'', funded by the German Excellence Strategy. We  also thank the German Ministry for Economic Affairs and Climate Action for funding in the project ``MAINN'' (funding ID 50OO2206). DRGS gratefully acknowledges support by the ANID BASAL projects ACE210002 and FB21003, as well as via the Millenium Nucleus NCN19-058 (TITANs). DRGS thanks for funding via Fondecyt Regular (project code 1201280).

%%%%%%%%%%%%%%%%%%%%%%%%%%%%%%%%%%%%%%%%%%%%%%%
\section*{Data Availability}
All data relevant for this article are included in the article itself. Potentially missing information can be requested from the first author via e-mail.
 
%The inclusion of a Data Availability Statement is a requirement for articles published in MNRAS. Data Availability Statements provide a standardised format for readers to understand the availability of data underlying the research results described in the article. The statement may refer to original data generated in the course of the study or to third-party data analysed in the article. The statement should describe and provide means of access, where possible, by linking to the data or providing the required accession numbers for the relevant databases or DOIs.



\input{mnras_template.bbl}

%\bibliographystyle{mnras}

%\begin{thebibliography}{99}
%\end{thebibliography}

%%%%%%%%%%%%%%%%%%%%%%%%%%%%%%%%%%%%%%%%%%%%%%%%%%

%%%%%%%%%%%%%%%%% APPENDICES %%%%%%%%%%%%%%%%%%%%%

%%%%%%%%%%%%%%%%%%%%%%%%%%%%%%%%%%%%%%%%%%%%%%%%%%

% Don't change these lines
\bsp	% typesetting comment
\label{lastpage}
\end{document}